\documentclass[runningheads]{llncs}
\usepackage{graphicx}
\usepackage{subcaption}
\usepackage{listings}[language=Python]
\usepackage{pgfplots,adjustbox}
\usepackage[title]{appendix}
\pgfplotsset{compat=1.15}
\usepackage[english]{babel}
\usepackage{makecell}
\usepackage{amsmath}
\usepackage{color}
\usepackage{algorithmic}
\usepackage{textcomp}
\usepackage{bmpsize}
\usepackage{xcolor, colortbl}
\usepackage{lipsum}
\usepackage{hyperref}
\usepackage[font={small}, justification=centering]{caption}
\usepackage{hyphenat}
\newcounter{nalg} % defines algorithm counter for chapter-level
\renewcommand{\thenalg}{\arabic{nalg}} %defines appearance of the algorithm counter
\DeclareCaptionLabelFormat{algocaption}{Algorithm \thenalg} % defines a new caption label as Algorithm x.y

\lstnewenvironment{algorithm}[1][] %defines the algorithm listing environment
{   
	\refstepcounter{nalg} %increments algorithm number
	\captionsetup{labelformat=algocaption,labelsep=colon} %defines the caption setup for: it ises label format as the declared caption label above and makes label and caption text to be separated by a ':'
	\lstset{ %this is the stype
		mathescape=true,
		frame=tB,
		basicstyle=\scriptsize, 
		keywordstyle=\color{black}\bfseries\em,
		keywords={,Inputs, Output, return, datatype, function, in, if, else, foreach, while, begin, end, },
		xleftmargin=0.0\textwidth,
		#1 % this is to add specific settings to an usage of this environment (for instnce, the caption and referable label)
	}
}
{}

\usepackage{url}
\begin{document}
\title{Anonymizing Machine Learning Models}
%
%\titlerunning{Abbreviated paper title}
% If the paper title is too long for the running head, you can set
% an abbreviated paper title here
%
\author{Abigail Goldsteen\inst{1} \and
Gilad Ezov\inst{1} \and
Ron Shmelkin\inst{1} \and
Micha Moffie\inst{1} \and
Ariel Farkash\inst{1} }
\authorrunning{A. Goldsteen et al.}
% First names are abbreviated in the running head.
% If there are more than two authors, 'et al.' is used.
%
\institute{IBM Research - Haifa, Haifa University Campus, Haifa, Israel 
\email{\{abigailt,ronsh,moffie,arielf\}@il.ibm.com, Gilad.Ezov@ibm.com}\\
\url{http://www.research.ibm.com/labs/haifa/} }
\maketitle              % typeset the header of the contribution
\begin{abstract}
  There is a known tension between the need to analyze personal data to drive business and privacy concerns. Many data protection regulations, including the EU General Data Protection Regulation (GDPR) and the California Consumer Protection Act (CCPA), set out strict restrictions and obligations on the collection and processing of personal data. Moreover, machine learning models themselves can be used to derive personal information, as demonstrated by recent membership and attribute inference attacks. Anonymized data, however, is exempt from the obligations set out in these regulations. It is therefore desirable to be able to create models that are anonymized, thus also exempting them from those obligations, in addition to providing better protection against attacks. 
  
  Learning on anonymized data typically results in significant degradation in accuracy. In this work, we propose a method that is able to achieve better model accuracy by using the knowledge encoded within the trained model, and guiding our anonymization process to minimize the impact on the model’s accuracy, a process we call \emph{accuracy-guided anonymization}. We demonstrate that by focusing on the model’s accuracy rather than generic information loss measures, our method outperforms state of the art k-anonymity methods in terms of the achieved utility, in particular with high values of k and large numbers of quasi-identifiers. 
  
  We also demonstrate that our approach has a similar, and sometimes even better ability to prevent membership inference attacks as approaches based on differential privacy, while averting some of their drawbacks such as complexity, performance overhead and model-specific implementations. In addition, since our approach does not rely on making modifications to the training algorithm, it can even work with ``black-box'' models where the data owner does not have full control over the training process, or within complex machine learning pipelines where it may be difficult to replace existing learning algorithms with new ones. This makes model-guided anonymization a legitimate substitute for such methods and a practical approach to creating privacy-preserving models. 
\end{abstract}
\section{Introduction}
The EU General Data Protection Regulation (GDPR)\footnote{\url{https://ec.europa.eu/info/law/law-topic/data-protection/data-protection-eu\_en}} sets out many restrictions on the processing of personal data. Similarly, the California Consumer Protection Act (CCPA)\footnote{\url{https://leginfo.legislature.ca.gov/faces/billTextClient.xhtml?bill\_id=201720180AB375}}, as well as the superseding California Privacy Rights Act (CPRA), set out several consumer rights in relation to the collection of personal information. Similar laws and regulations are being enacted in additional states and countries around the world. Adhering to these regulations can be a complex and costly task. 

However, these regulations specifically exempt anonymized data. Recital 26 of GDPR states that the principles of data protection should not apply to anonymous information, which does not relate to an identified or identifiable natural person, or to personal data rendered anonymous in such a manner that the data subject is no longer identifiable. CCPA, in its article 1798.145 (a), affirms that the obligations imposed by the title shall not restrict a business’s ability to collect, use, retain, sell, or disclose consumer information that is deidentified or aggregate consumer information. ``Deidentified'' in this context means information that cannot reasonably identify, relate to, describe, be capable of being associated with, or be linked, directly or indirectly, to a particular consumer. It is therefore very attractive for data collectors to be able to perform their processing tasks on anonymized data.

Many data processing tasks nowadays involve machine learning (ML). In recent years, several attacks have been developed that are able to infer sensitive information from trained models. Examples include \emph{membership inference attacks}, where one can deduce whether a specific individual was part of the training set or not; and \emph{model inversion attacks} or \emph{attribute inference attacks}, where certain sensitive features may be inferred about individuals who participated in training a model. This has led to the conclusion that machine learning models themselves should, in some cases, be considered personal information \cite{Veale18}, \cite{Kazim21}, and therefore subject to GDPR and similar laws. The 2020 study of the European Parliamentary Research Service (EPRS) on the impact of GDPR on artificial intelligence\footnote{https://www.europarl.europa.eu/RegData/etudes/STUD/2020/641530/EPRS\_STU(2020)641530\_EN.pdf} found that, although AI is not explicitly mentioned in the GPDR, many provisions in the GDPR are relevant to AI. The authors note that for an item to be linked to a person, it is not necessary to identify the data subject with absolute certainty; a degree of probability may be sufficient. They propose to address this by ensuring that data is de-identified in ways that make it more difficult to re-identify the data subject. 

It is therefore desirable to be able to anonymize the AI models themselves, i.e., ensure that the personal information of a specific individual that participated in the training set cannot be re-identified. 
Most existing work on protecting the privacy of machine learning training data, including differential privacy, typically requires making changes to the learning algorithms themselves, as they incorporate the perturbation into the model training process \cite{dp}, \cite{pate1}. They are thus not suitable for scenarios in which the learning process is performed by a third party and is not under the control of the organization that owns (and wants to anonymize) the private data. They may also be extremely difficult to adopt in organizations that employ many different ML models of different types. Moreover, when applying this type of method, any effort already invested in model selection and hyperparameter tuning may need to be redone, since the learning algorithm itself is replaced.
In this paper, we present a practical solution for anonymizing machine learning models that is completely agnostic to the type of model trained.

Akin to some previous efforts, the method is based on applying k-anonymity to the training data and then training the model on the anonymized dataset.  Past attempts at training machine learning models on anonymized data have resulted in very poor accuracy \cite{perturbed}, \cite{anonML}, and it is therefore typically not employed in practice. However, our anonymization method is guided by the specific machine learning model that is going to be trained on the data. We make use of the knowledge encoded within the model to produce an anonymization that is highly tailored. We call this method \emph{accuracy-guided anonymization}. Once the training dataset is anonymized in this manner, retraining of the model is performed to yield an anonymized model. 

 We demonstrate that this approach outperforms state of the art anonymization techniques in terms of the achieved utility, as measured by the resulting model's accuracy. We also show that our proposed approach can preserve an acceptable level of accuracy even with fairly high values of k and larger numbers of quasi-identifier attributes, making anonymous machine learning a feasible option for many enterprises. 

We also tested the effectiveness of our method as a mitigation against membership inference attacks and compared it against existing implementations of differentially private models. We found that the results achieved using our anonymization were comparable to, sometimes even slightly better than, differential privacy in terms of the achieved model accuracy for the same level of protection against attacks. This shows that model-guided anonymization can, in some cases, be a legitimate substitute for such methods, while averting some of their inherent drawbacks such as complexity, performance overhead and the need for different implementations for each model type.

Our approach is generic and can be applied to any type of machine learning model. Since it does not rely on making modifications to the training process, it can be applied in a wide variety of use cases, including integration within existing ML pipelines, or in combination with machine learning as a service (ML-as-a-service or MLaaS in short). This setting is particularly useful for organizations that do not possess the required computing resources to perform their training tasks locally. It can even be applided to existing models, reusing the same architecture and hyperparameters and requiring only retraining the model.

Similar to classic k-anonymity methods, our method can only be applied to structured data, including numeric, discrete and categorical features. It also does not work well for very high-dimensional data, or data with a very high degree of uniqueness.

In the remainder of the paper we present the most relevant related work in Section \ref{sec:rel}, describe the details of our method in Section \ref{sec:anon}, present experimental results in Section \ref{sec:res} and finally conclude in Section \ref{sec:con}.

\section{Related Work}
\label{sec:rel}
\subsection{K-anonymity}
K-anonymity was proposed by L. Sweeney \cite{Sweeney2002} to address the problem of releasing personal data while preserving individual privacy. It is a method to reduce the likelihood of any single person being identified when the dataset is linked with external data sources. The approach is based on generalizing attributes, and possibly deleting records, until each record becomes indistinguishable from at least $k-1$ others. A generalization of a numeric attribute typically consists of a range of consecutive values, whereas the generalization of a categorical attribute consists of a sub-group of categories. Generalization is applied only to attributes that can be used in combination, or when linked with other data sources, to enable re-identifying individuals. Such attributes are called quasi-identifiers (QI). 

%This initial work was followed by many extensions and improvements. Since achieving optimal k-anonymity is NP-hard, many polynomial heuristics have been proposed \cite{Incognito}, \cite{mondrian}, \cite{hilbert}. Researchers have also focused on maximizing the utility of the anonymized result \cite{iloss}. 

There has been some criticism of anonymization methods in general and k-anonymity in particular, mostly revolving around possible re-identification even after a dataset has been anonymized. Cases such as the Netflix recommendation contest dataset \cite{netflix} have been used to justify the need for new, more robust methods. However, a deeper analysis of these cases reveals that this typically occurs when poor anonymization techniques were applied \cite{avoid} or when the chosen list of quasi-identifiers was not exhaustive, leaving some attributes that could potentially be linked to external datasets unaccounted for. When correctly applied, the re-identification risk in a k-anonymized dataset is at most $1/k$ \cite{k}.

Another shortcoming of basic k-anonymity is that even though it prevents identity disclosure (re-identification), it may fail to protect against attribute disclosure if all (or most of) the k records within a similarity group share the same value of a sensitive attribute \cite{critique}. For example, in the case of health records, if all k records within a group share the same disease, knowing that an individual belongs to that group automatically discloses their medical condition. To this end, extensions such as \textit{l}-diversity \cite{l-diversity} and \textit{t}-closeness \cite{t-closeness} were developed.

%Xiao et. al. \cite{personalized} present the concept of personalized anonymity, and describe a framework that performs the minimum generalization for satisfying each individual's personal privacy requirements, thus retaining the largest amount of information possible from the microdata. Hajian et. al. \cite{generalization} present a generalization-based approach to simultaneously offer privacy preservation and discrimination prevention. However, this solution is mainly focused on acheiving non-discrimination and not on tailoring or improving the anonymity method itself.

\subsection{Protecting Machine Learning Training Sets}
Several attacks have been able to reveal either whether an individual was part of the training set (membership inference attacks) or infer certain possibly sensitive properties of the training data (attribute inference attacks) \cite{warfarin}, \cite{confidence}, \cite{membership}, \cite{leaks}. As a result, a lot of work has focused on protecting the privacy of datasets used to train machine learning models. 

One straightforward approach to applying the concept of k-anonymity to machine learning is to perform anonymization on the training data prior to training the model. When these methods are applied without considering the intended use of the data, they tend to yield poor accuracy results \cite{perturbed}, \cite{anonML}. Iyengar \cite{constraints} and LeFevre et al. \cite{workload} were the first to propose \emph{tailored anonymization} that takes into consideration the intended use of the data, e.g., training a machine learning model. They do not, however, tailor their methods to a specific machine learning model, since their goal is to publish the anonymized dataset. Rather they use either a generic ``classification metric'' \cite{constraints} or simply use the label that is the target of classification \cite{workload}. 
 In our work, we use an initial, trained machine learning model as a starting point for the anonymization process, using the model's predictions on the data as our similarity measure, as described in detail in Section \ref{sec:anon}. For different target models, a different version of the anonymized dataset will be created, thus yielding an anonymization that is truly tailored to the model.

In addition, the approach described in \cite{workload} for mapping the resulting regions to a set of features on which to train the model forces them to apply the same recoding mechanism to any test data before the model can be applied to it. This may make applying the method more difficult in practice. In our method, test data can be used as is in the anonymized model, without having to apply any special transformation to it. 

More recently, approaches based on \emph{differential privacy} (DP), have been used to add noise during training to protect training sets and counter membership attacks \cite{dp}, \cite{rf}, \cite{lib}. The idea behind this approach is to reduce the effect of any single individual on the model's outcome. These techniques tend to be highly tailored to specific machine learning algorithms, including internal implementation details such as the choice of loss function or optimization algorithm.

Another significant challenge when using DP in practice is the choice of privacy budget ($\epsilon$ value). A recent survey of techniques based on differential privacy \cite{diffpriv} analyzed various variants of DP that differ in their analysis of the cumulative privacy loss, and thereby in the amount of noise added during training to satisfy a particular privacy budget. They found that the level of privacy leakage (measured by the success of inference attacks) for each variant accords with the actual amount of noise added to the model, and not necessarily with the absolute $\epsilon$ value, indicating that $\epsilon$ alone is not a good indicator for the degree of privacy achieved. Their main finding is that current mechanisms for differentially private machine learning rarely offer acceptable utility-privacy trade-offs for complex learning tasks. 
Moreover, Bagdasaryan et al. \cite{disparate} and Melis et al. \cite{collab} showed that differentially private training may fail to converge in some cases.

PATE \cite{pate1} transfers to a ``student'' model the knowledge of an ensemble of ``teacher'' models, with privacy guaranteed by noisy aggregation of teachers' answers. However, this approach assumes the availability of a public or non-sensitive dataset with a similar distribution to the private dataset on which the student model can be trained. This may be very difficult to achieve in practice, especially when dealing with highly sensitive and unique datasets collected exclusively by governmental agencies, pharmaceutical companies, etc.

The most significant difference between approaches based on differential privacy or ensemble aggregation and our approach is that they require complete control over the learning process. The learning algorithms themselves must be modified to guarantee the desired privacy properties. These approaches cannot be applied at all in cases where the learning process occurs outside the hands of the private data owner, such as with ML-as-a-service. Since our method does not rely on making modifications to the training process, it is much better suited for such (and similar) scenarios.

\section{Model-guided anonymization}
\label{sec:anon}
In this section we describe in more detail the proposed approach for machine learning model anonymization. Our implementation is based on k-anonymity, but could possibly be extended to cover additional guarantees such as l-diversity and t-closeness.
Many typical k-anonymity algorithms rely on finding groups of k (or more) similar records that can be mapped together to the same generalized value (on the quasi-identifier attributes), thus fulfilling the requirement that these k records are indistinguishable from each other. Some do this by iteratively choosing a feature and splitting the domain of the feature along a chosen split value. The resulting splits create a partitionning of the domain into groups of k or more records, which serve as the basis for the generalization.

The Median Mondrian method \cite{mondrian} chooses at each phase the split attribute with the widest normalized range and uses the median value as the split value. The Greedy algorithm \cite{iloss} chooses the candidate split based on minimal information loss. In our case, we replace these criteria with minimal accuracy loss in the target model. 

\begin{figure*}
	\centering
	\includegraphics[width=10cm]{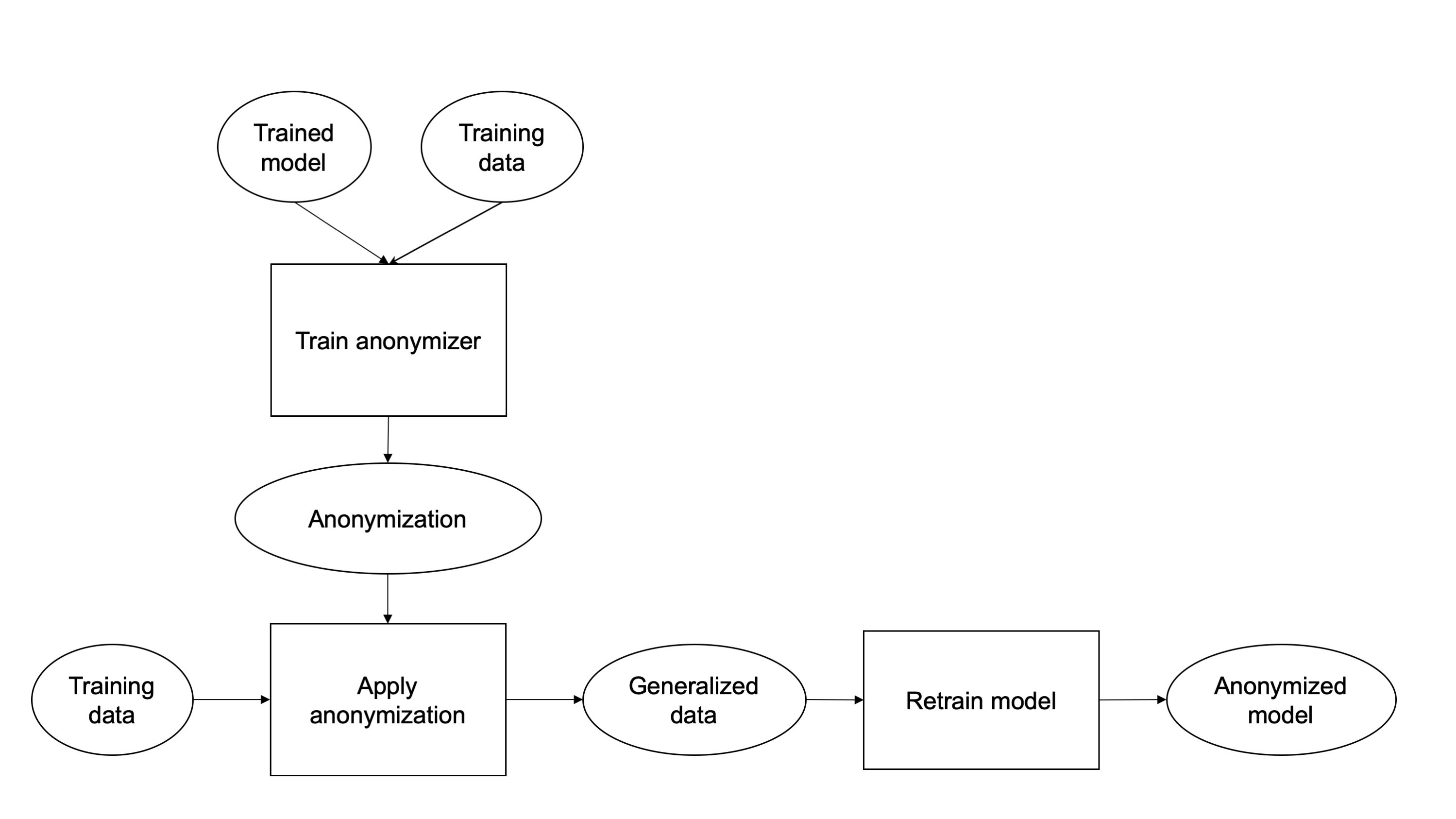}
	\caption{Complete anonymization process}
	\label{fig:process}
\end{figure*}

The overall process of \emph{accuracy-guided anonymization} is depicted in Figure \ref{fig:process}. The process starts with an initial model trained on the raw training data. This is the model whose accuracy we will try to preserve. To achieve this, we use the training data with the initial model's predictions on that data as input to the anonymization process. 

Next we train an \emph{anonymizer model} on the training data, using as labels the original model's predictions. That is, we train a new model to learn the target model's  ``decision boundaries''. This is similar to the student-teacher training or knowledge distillation concept often employed as a means for model compression \cite{dist}.
In use cases where the full model training is performed by a third (possibly untrusted) party, the initial model used to generate these predictions may be a simple, representative model, trained on a small subset of the data or a pre-trained model performing a similar classification task as the target model. If no pre-existing or minimally trained model is available, the target (class) labels may be used instead of model predictions.

For the anonymizer model we employ a decision tree model. 
We set the minimum number of samples required to be in each leaf node of the tree to k, and then use the leaves of the tree as the basis for generalization. The training samples mapped to each leaf constitute the group of records that are generalized to the same value. Since each leaf node contains at least k samples from the training set, and we generalize all of those samples in the same manner, they will be indistinguishable from each other, thus satisfying k-anonymity.

One common approach to generalize features for a group of records is to use a range (for numerical features) or sub-group (for categorical features) that covers all records in the group. For example, if we have three records with the age values 31, 32 and 34, these can be generalized to the age range 31-34. Similarly, if those three records have the occupation values ``nurse'', ``teacher'' and ``engineer'', these can be generalized to the group [``nurse'',``teacher'',``engineer'']. However, since we want to be able to retrain our ML model on the anonymized dataset, we need to map these ranges and sub-groups to numerical features on which the model can be trained.

In \cite{workload}, the d-dimensional hyper-rectangle of each partition is represented as points in a 2-dimensional space. We adopted a different approach, where we map each sample to a concrete, representative point in the same domain as the original features. Having the anonymized records share the same domain as the original data enables using the model directly on any newly collected test data, without having to apply any special recoding dictated by the anonymization process. In other words, the rest of the ML lifecycle can remain unchanged. 

There are several choices for mapping the data points in each leaf to a representative value. We opted to use an actual value that falls in the cluster as the representative point for that cluster, as we found that it entails higher prediction accuracy. We chose to use the point closest to the median of the cluster from the points with the majority label in that cluster. It is also possible to choose an actual value from each range/group separately. By mapping all records belonging to the same leaf node to the same concrete value, we still satisfy the k-anonymity requirement. 

Algorithm \ref{code} describes the anonymization part of the algorithm. 

\vspace{-1em}

\begin{center}
	\begin{algorithm}[caption={Anonymization algorithm}, label={code}]
Inputs: training data $X$ labeled with original model's predictions $y$, 
	list of quasi-identifiers QI, required $k$ value
Output: anonymized training data $\overline{\rm X}$
		
Separate $X$ into $X_{qi}$ (only QI features) and $X_{rest}$ (rest of features)
Train decision tree $T$ on ($X_{qi}$, $y$) with min_samples_leaf=$k$
foreach leaf node $l$ in $T$:
	$$S $\gets$ samples in $l$
	$$m $\gets$ median($S$)
	$s$ $\gets$ sample closest to $m$ (using euclidean distance)
$S$' $\gets$ replace all samples in $S$ with $s$
$X$' $\gets$ $\cup$ $S$'
Re-attach $X$' with $X_{rest}$ $\to$ $\overline{\rm X}$
	\end{algorithm}
\end{center}

Once each of the original records is mapped to its new, representative value, based on the leaf to which it belongs, the dataset is k-anonymized.
Finally, we retrain the model on the anonymized data, resulting in an anonymized model that is no loger considered personal information (e.g., under Recital 26 of GDPR) and can be stored and shared freely. 

The described anonymization process is typically performed after applying feature selection. The motivation for this is that if there are features that are not needed at all, they should simply be removed from consideration rather than undergo anonymization. 
However, there are also advantages to applying our method before feature selection. When performing feature selection ``blindly'', without knowing which features are considered quasi-identifiers, the process may inadvertently select quasi-identiers as important features, even if other, non-identifying features could have been used instead. In these cases, it may be better to perform feature selection after anonymization, allowing features that have not been anonymized to be prioritized.
In Section \ref{sec:res} we show results both with and without applying feature selection prior to the anonymization.

The code for our accuracy-guided anonymization method can be found on github: https://github.com/IBM/ai-privacy-toolkit.

\section{Results}
\label{sec:res}
Our evaluation method consists of the following steps. First, we select a dataset, train one or more original models on it (either applying feature selection or not), and measure their accuracy. We consider the resulting model and its accuracy as our baseline. We then perform the anonymization process described in Section \ref{sec:anon} and retrain the model on the anonymized data. Finally, we measure the accuracy of the new model on a hold-out test set. 

We evaluated our method using two openly available datasets: Adult\footnote{https://archive.ics.uci.edu/ml/datasets/adult}, an excerpt of the 1994 U.S. Census database, which is widely used to evaluate anonymization and generalization techniques; and Loan\footnote{https://www.lendingclub.com/info/download-data.action}, an excerpt of the Lending Club loan data from 2015. This second dataset was chosen to demonstrate that our approach works for data with a larger number of attributes.
The characteristics of each dataset are presented in Appendix \ref{app:QI}. 

Both datasets were used for binary classification tasks. 
For the Adult dataset we trained neural network (NN) (86.5), XGBoost (XGB) (86.4), decision tree (DT) (88.78) and random forest (RF) (89.92) models (the test accuracy of each model is stated in parentheses). The last two were chosen because of their use in \cite{workload} and \cite{constraints}. The neural network was composed of one hidden layer with 100 neurons, relu activation and adam optimizer, and a constant learning rate of 0.001. The random forest was composed of 100 trees using Gini impurity as the split criterion. Except for neural networks, in all other cases we performed feature selection prior to anonymization. 
We also chose several different sets of quasi-identifiers: we either used all 12 attributes, 10 attributes or 8 attributes (as detailed in Appendix \ref{app:QI}). The list of 8 attributes was chosen to be similar to the one used in \cite{constraints}. Due to space constraints we show results for select runs. The remainder of the results follow similar patterns. 

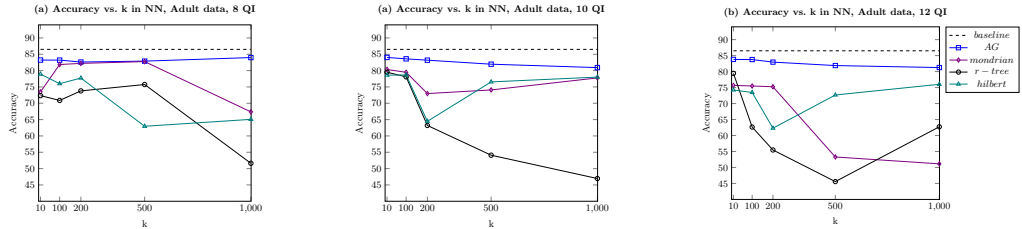
\begin{figure*}
	\begin{subfigure}{.33\textwidth}
		\resizebox{.75\columnwidth}{!}{%
		\begin{tikzpicture}[baseline,trim axis left]
		\begin{axis}[legend style={font=\small}, title style={font=\small}, xlabel=k, ylabel=Accuracy, xmin=1,xmax=1001, ymin=40, ymax=94, ytick={45, 50, 55, 60, 65, 70, 75, 80, 85, 90}, xtick={10, 100, 200, 500, 1000},  title={\bfseries (a) Accuracy vs. k in NN, Adult data, 8 QI}, 
		legend pos=outer north east]
%		legend pos=south west, at={(-1,0)},anchor=west, y label style={at={(0.07,0.4)},anchor=west}]
		\addplot [mark=none, black, dashed] coordinates { (10,86.5) (1000,86.5) }; 
%		\addlegendentry{$baseline$}
		\addplot [color=blue, mark=square] coordinates {( 10, 83.23 )	( 100, 83.21 )	( 200, 82.62 )	( 500, 82.9 )	( 1000, 83.99 ) }; 
%		\addlegendentry{$AG$}
		\addplot [color=violet, mark=diamond] coordinates {( 10, 73.49 )	( 100, 81.86 )	( 200, 82.21 )	( 500, 82.76 )	( 1000, 67.36 )}; 
%		\addlegendentry{$mondrian$}
		\addplot [color=black, mark=o] coordinates { ( 10, 72.34 )	( 100, 70.87 )	( 200, 73.79 )	( 500, 75.74 )	( 1000, 51.62 )};
%		\addlegendentry{$r-tree$}
		\addplot [color=teal, mark=triangle] coordinates { ( 10, 78.91) ( 100, 76 )	( 200,77.72 )	( 500, 62.96 )	( 1000, 65.07 )};
%		\addlegendentry{$hilbert$}
		\end{axis}
		\end{tikzpicture}}		
	\end{subfigure}%	
	\hspace*{0.2in}
	\begin{subfigure}{.33\textwidth}
		\resizebox{.75\columnwidth}{!}{%
			\begin{tikzpicture}[baseline,trim axis left]
			\begin{axis}[legend style={font=\small}, title style={font=\small}, xlabel=k, ylabel=Accuracy, xmin=1,xmax=1001, ymin=40, ymax=94, ytick={45, 50, 55, 60, 65, 70, 75, 80, 85, 90}, xtick={10, 100, 200, 500, 1000},  title={\bfseries (a) Accuracy vs. k in NN, Adult data, 10 QI}, 
			legend pos=outer north east]
			%		legend pos=south west, at={(-1,0)},anchor=west, y label style={at={(0.07,0.4)},anchor=west}]
			\addplot [mark=none, black, dashed] coordinates { (10,86.5) (1000,86.5) }; 
%			\addlegendentry{$baseline$}
			\addplot [color=blue, mark=square] coordinates {( 10, 84.03 )	( 100, 83.6 )	( 200, 83.19 )	( 500, 81.96 )	( 1000, 80.92 ) }; 
%			\addlegendentry{$AG$}
			\addplot [color=violet, mark=diamond] coordinates {( 10, 80.34 )	( 100, 79.48 )	( 200,72.97 )	( 500, 74.08 )	( 1000, 77.76 )}; 
%			\addlegendentry{$mondrian$}
			\addplot [color=black, mark=o] coordinates { ( 10, 79.5 )	( 100, 78.09 )	( 200, 63.21 )	( 500, 54.1 )	( 1000, 46.96 )};
%			\addlegendentry{$r-tree$}
			\addplot [color=teal, mark=triangle] coordinates { ( 10, 78.6 ) ( 100, 78.66 )	( 200,64.48 )	( 500, 76.54 )	( 1000, 78.01 )};
%			\addlegendentry{$hilbert$}
			\end{axis}
			\end{tikzpicture}}
	\end{subfigure}%	
	\hspace*{0.2in}
	\begin{subfigure}{.33\textwidth}
		\resizebox{.95\columnwidth}{!}{%
			\begin{tikzpicture}[baseline,trim axis left]
			\begin{axis}[legend style={font=\small}, title style={font=\small}, xlabel=k, ylabel=Accuracy, xmin=1,xmax=1001, ymin=40, ymax=94, ytick={45, 50, 55, 60, 65, 70, 75, 80, 85, 90}, xtick={10, 100, 200, 500, 1000},  title={\bfseries (b) Accuracy vs. k in NN, Adult data, 12 QI}, 
			legend pos=outer north east]
			%		legend pos=south west, at={(-1,0)},anchor=west, y label style={at={(0.07,0.4)},anchor=west}]
			\addplot [mark=none, black, dashed] coordinates { (10,86.5) (1000,86.5) }; 
			\addlegendentry{$baseline$}
			\addplot [color=blue, mark=square] coordinates {( 10, 83.8 ) ( 100, 83.76 )	( 200, 82.96 )	( 500, 81.88 )	( 1000, 81.26 ) }; 
			\addlegendentry{$AG$}
			\addplot [color=violet, mark=diamond] coordinates {( 10, 75.68 ) ( 100, 75.49 )	( 200, 75.25 )	( 500, 53.28 )	( 1000, 51.13 )}; 
			\addlegendentry{$mondrian$}
			\addplot [color=black, mark=o] coordinates { ( 10, 79.44 ) ( 100, 62.64 )	( 200, 55.47 )	( 500, 45.58 )	( 1000, 62.72 )};
			\addlegendentry{$r-tree$}
			\addplot [color=teal, mark=triangle] coordinates { ( 10, 74.24 ) ( 100, 73.46 )	( 200, 62.29 )	( 500, 72.67 )	( 1000, 76.02 )};
			\addlegendentry{$hilbert$}
			\end{axis}
			\end{tikzpicture}}
	\end{subfigure}%
	\caption{Adult data with NN model, different sets of QI}
	\label{fig:a12_graph}
\end{figure*}

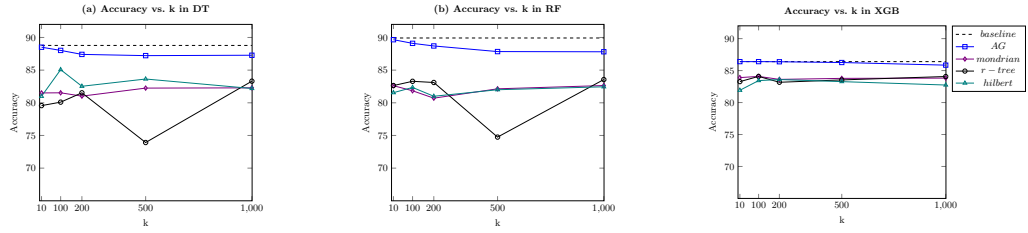
\begin{figure*}
	\begin{subfigure}{.33\textwidth}
		\resizebox{.75\columnwidth}{!}{%
			\begin{tikzpicture}[baseline,trim axis left]
			\begin{axis}[legend style={font=\small}, title style={font=\small}, xlabel=k, ylabel=Accuracy, xmin=1,xmax=1001, ymin=65, ymax=92, ytick={70, 75, 80, 85, 90}, xtick={10, 100, 200, 500, 1000},  title={\bfseries (a) Accuracy vs. k in DT}, 
			legend pos=south west]
%			legend pos=south west, at={(-1,0)},anchor=west, y label style={at={(0.07,0.4)},anchor=west}]
%			legend pos=outer north east, at={(-1,0)},anchor=west, y label style={at={(0.07,0.4)},anchor=west}]
			\addplot [mark=none, black, dashed] coordinates { (10,88.782) (1000,88.78) }; 
%			\addlegendentry{$baseline$}
			\addplot [color=blue, mark=square] coordinates {( 10, 88.51 )	( 100, 88.02 )	( 200, 87.41 )	( 500, 87.22 )	( 1000, 87.28 ) }; 
%			\addlegendentry{$AG$}
			\addplot [color=violet, mark=diamond] coordinates {( 10, 81.51 )	( 100, 81.53 )	( 200, 81.02 )	( 500, 82.25 )	( 1000, 82.29 )}; 
%			\addlegendentry{$mondrian$}
			\addplot [color=black, mark=o] coordinates { ( 10,79.57 )	( 100, 80.1 )	( 200, 81.53 )	( 500, 73.92)	( 1000, 83.31 )};
%			\addlegendentry{$r-tree$}
			\addplot [color=teal, mark=triangle] coordinates { ( 10, 81 ) ( 100, 85.07 )	( 200,82.53 )	( 500, 83.64 )	( 1000, 82.19 )};
%			\addlegendentry{$hilbert$}
			\end{axis}
			\end{tikzpicture}}
	\end{subfigure}
	\hspace*{0.2in}
	\begin{subfigure}{.33\textwidth}
		\resizebox{.75\columnwidth}{!}{%
			%		\resizebox{.88\columnwidth}{!}{%
			\begin{tikzpicture}[baseline,trim axis left]
			\begin{axis}[legend style={font=\small}, title style={font=\small}, xlabel=k, ylabel=Accuracy, xmin=1,xmax=1001, ymin=65, ymax=92, ytick={70, 75, 80, 85, 90}, xtick={10, 100, 200, 500, 1000},  title={\bfseries (b) Accuracy vs. k in RF}, 
			legend pos=south west]
%			legend pos=outer north east, at={(-1,0)},anchor=west, y label style={at={(0.07,0.4)},anchor=west}]
			\addplot [mark=none, black, dashed] coordinates { (10,89.92) (1000,89.92) }; 
%			\addlegendentry{$baseline$}
			\addplot [color=blue, mark=square] coordinates {( 10, 89.66 )	( 100, 89.1 )	( 200, 88.7 )	( 500, 87.84 )	( 1000, 87.81 ) }; 
%			\addlegendentry{$AG$}
			\addplot [color=violet, mark=diamond] coordinates {( 10, 82.62 )	( 100, 81.84 )	( 200, 80.71 )	( 500, 82.14 )	( 1000, 82.64) }; 
%			\addlegendentry{$mondrian$}
			\addplot [color=black, mark=o] coordinates { ( 10, 82.64 )	( 100, 83.29 )	( 200, 83.11 )	( 500, 74.75 )	( 1000, 83.56) };
%			\addlegendentry{$r-tree$}
			\addplot [color=teal, mark=triangle] coordinates { ( 10, 81.55 ) ( 100, 82.35 )	( 200,81 )	( 500, 82 )	( 1000, 82.45 )};
%			\addlegendentry{$hilbert$}
			\end{axis}
			\end{tikzpicture}}
	 	\end{subfigure}%	
		\hspace*{0.2in}
		\begin{subfigure}{.33\textwidth}
			\resizebox{.95\columnwidth}{!}{%
			\begin{tikzpicture}[baseline,trim axis left]
			\begin{axis}[legend style={font=\small}, title style={font=\small}, xlabel=k, ylabel=Accuracy, xmin=1,xmax=1001, ymin=65, ymax=92, ytick={70, 75, 80, 85, 90}, xtick={10, 100, 200, 500, 1000},  title={\bfseries Accuracy vs. k in XGB}, 
			legend pos=outer north east]
			\addplot [mark=none, black, dashed] coordinates { (10,86.4) (1000,86.4) }; 
			\addlegendentry{$baseline$}
			\addplot [color=blue, mark=square] coordinates {( 10, 86.4 )	( 100, 86.4 )	( 200, 86.38 )	( 500, 86.26 )	( 1000, 85.85 ) }; 
			\addlegendentry{$AG$}
			\addplot [color=violet, mark=diamond] coordinates {( 10, 83.9 )	( 100, 84.09)	( 200, 83.62 )	( 500, 83.78)	( 1000, 83.82 )}; 
			\addlegendentry{$mondrian$}
			\addplot [color=black, mark=o] coordinates { ( 10, 83.27 )	( 100, 84.09 )	( 200, 83.17 )	( 500, 83.54 )	( 1000, 84.07 )};
			\addlegendentry{$r-tree$}
			\addplot [color=teal, mark=triangle] coordinates { ( 10, 81.92 ) ( 100, 83.43 )	( 200,83.6 )	( 500, 83.27 )	( 1000, 82.74 )};
			\addlegendentry{$hilbert$}
			\end{axis}
			\end{tikzpicture}}
	\end{subfigure}
	\caption{Adult data with 8 QI+feature selection, different models}
	\label{fig:a8_fs_graph}
\end{figure*}

The data was divided into three subsets, used for: (1) training the original model (40\%), (2) training the anonymizer decision tree model (40\%), and (3) validation (20\%). The 40\% data that was used to train the original model is also the dataset that is eventually anonymized and then used to retrain the model. The validation set was used to measure the accuracy of the new, anonymized model.
Each run of our accuracy-guided anonymization (AG) is compared with three ``regular'' anonymization methods: the Median Mondrain method \cite{mondrian}, the Hilbert-curve based method \cite{hilbert} and the R+ tree based method \cite{rtree}. 
The results of these runs can be found in Figures \ref{fig:a12_graph} and \ref{fig:a8_fs_graph}. 

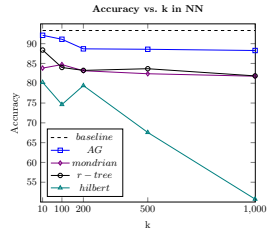
\begin{figure}
	\hspace*{1.8in}
	\resizebox{.25\columnwidth}{!}{%	
		\begin{tikzpicture}[baseline,trim axis left]
		\begin{axis}[legend style={font=\small}, title style={font=\small}, xlabel=k, ylabel=Accuracy, xmin=1,xmax=1001, ymin=50, ymax=95, ytick={55, 60, 65, 70, 75, 80, 85, 90}, xtick={10, 100, 200, 500, 1000},  title={\bfseries Accuracy vs. k in NN}, 
		legend pos=south west]
		\addplot [mark=none, black, dashed] coordinates { (10,93.32) (1000,93.32) }; 
		\addlegendentry{$baseline$}
		\addplot [color=blue, mark=square] coordinates {( 10, 92.11 )	( 100, 91.11 )	( 200, 88.69 )	( 500, 88.58 )	( 1000, 88.27 ) }; 
		\addlegendentry{$AG$}
		\addplot [color=violet, mark=diamond] coordinates {( 10, 83.83 )	( 100, 84.71 )	( 200, 83.16 )	( 500, 82.4 )	( 1000, 81.78 ) }; 
		\addlegendentry{$mondrian$}
		\addplot [color=black, mark=o] coordinates { ( 10, 88.39 )	( 100, 83.98 )	( 200, 83.25 )	( 500, 83.67 )	( 1000, 81.86 ) };
		\addlegendentry{$r-tree$}
		\addplot [color=teal, mark=triangle] coordinates { ( 10, 80.23 ) ( 100, 74.63 )	( 200,79.41 )	( 500, 67.57 )	( 1000, 50.74 )};
		\addlegendentry{$hilbert$}
		\end{axis}
		\end{tikzpicture}}
	\caption{Loan data with 18 QI, Neural Network model}
	\label{fig:l18_graph}
\end{figure}

For the Loan dataset we trained a neural network model (test accuracy: 93.32) with the same architecture and hyperparameters as the one used for Adult. For this dataset we used 18 attributes as QI (detailed in Appendix \ref{app:QI}). The results are presented in Figure \ref{fig:l18_graph}.

%Adult NN

% Adult other models

% Loan:

Next we compare our method with one of the workload-aware approaches, namely \cite{workload}. Since their method is basically equivalent to training a decision tree on the data with the true labels using an entropy-based split criterion, we employed a regular decision tree model, similar to the one used for model-guided training, but using the true labels instead of the target model's classifications. Note that we did not use the method they described for mapping the resulting regions into 2d points, but rather employed the same method that we used for choosing representative values for each cluster.

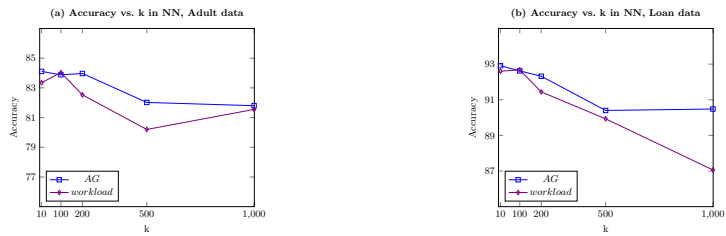
\begin{figure}
	\hspace*{0.6in}
	\begin{subfigure}[b]{.5\textwidth}
		\resizebox{.5\columnwidth}{!}{%
			\begin{tikzpicture}[baseline,trim axis left]
			\begin{axis}[legend style={font=\small}, title style={font=\small}, xlabel=k, ylabel=Accuracy, xmin=1,xmax=1001, ymin=75, ymax=87, ytick={77, 79, 81, 83, 85}, xtick={10, 100, 200, 500, 1000},  title={\bfseries (a) Accuracy vs. k in NN, Adult data}, 
			legend pos=south west]
%			\addplot [mark=none, black, dashed] coordinates { (10,86.5) (1000,86.5) }; 
%			\addlegendentry{$baseline$}
			\addplot [color=blue, mark=square] coordinates {( 10, 84.11 )	( 100, 83.88 )	( 200, 83.97 )	( 500, 82.02)	( 1000, 81.8 )  }; 
			\addlegendentry{$AG$}		
			\addplot [color=violet, mark=diamond] coordinates {( 10, 83.35 )	( 100, 84.03 )	( 200, 82.53 )	( 500, 80.2)	( 1000, 81.55 ) }; 
			\addlegendentry{$workload$}
			\end{axis}
			\end{tikzpicture}}
	\end{subfigure}%
	\begin{subfigure}[b]{.5\textwidth}
		\resizebox{.5\columnwidth}{!}{%
			\begin{tikzpicture}[baseline,trim axis left]
			\begin{axis}[legend style={font=\small}, title style={font=\small}, xlabel=k, ylabel=Accuracy, xmin=1,xmax=1001, ymin=85, ymax=95, ytick={87, 89, 91, 93}, xtick={10, 100, 200, 500, 1000},  title={\bfseries (b) Accuracy vs. k in NN, Loan data}, 
			legend pos=south west]
%			\addplot [mark=none, black, dashed] coordinates { (10,93.32) (1000,93.32) }; 
%			\addlegendentry{$baseline$}
			\addplot [color=blue, mark=square] coordinates {( 10, 92.91 )	( 100, 92.62 )	( 200, 92.32 )	( 500, 90.4)	( 1000, 90.48 )  }; 
			\addlegendentry{$AG$}		
			\addplot [color=violet, mark=diamond] coordinates {( 10, 92.6 )	( 100, 92.68 )	( 200, 91.44 )	( 500, 89.93)	( 1000, 87.06 ) }; 
			\addlegendentry{$workload$}
			\end{axis}
			\end{tikzpicture}}
	\end{subfigure}%	
	\caption{Neural Network model, AG vs. workload-aware anonymization}
	\label{fig:label_graph}
\end{figure}

For relatively simple models such as decision tree and random forest, the results were very similar between the two methods. This makes sense, especially for the decision tree, since the anonymization model and target model are equivalent. However, when moving to more complex models such as neural networks, a slight advantage of our method can be seen, as reflected in Figure \ref{fig:label_graph}. 

%\vspace{-3em}

\subsection{Discussion}
\label{sub:dis}
Figures \ref{fig:a12_graph} - \ref{fig:l18_graph} compare the results of our accuracy-guided anonymization method with three ``classic'' anonymization methods: Mondrian, Hilbert and R+ tree. These methods are not tailored for any specific analysis on the resulting dataset. It is clear that in all cases, our method outperforms these. 

Figure \ref{fig:label_graph} shows a comparison with workload-aware anonymization \cite{workload} that uses the label class value as an impurity measure when partitioning the data. Here, better accuracy was achieved mainly for more complex models such as neural networks. It is possible that in these cases, learning the ``decision boundaries'' of an existing model is more effective than learning the task from scratch (based on the true labels). This technique has proven useful to help distill the knowledge of a large, complex (teacher) model into smaller, simpler (student) models \cite{dist}. 
We therefore believe that our method is more advantageous when applied to more modern and complex models. It could be interesting to also compare the two methods of mapping clusters to features, but we leave this for future work.

It is noteworthy to point out that in many previous works on anonymization, results were demonstrated on relatively small values of k (up to 100) and small sets of quasi-identifiers (typically 8-10). In our evaluation, we demonstrated results for very high k values (up to 1000) and QI sets of up to 18 attributes, which, as far as we know, has not been attempted in the past. 
On one hand, as previously mentioned, it is critical to include as many QI attributes as possible to prevent re-identification. On the other hand, the more attributes are considered QI, the more generalization must be performed on the data and the higher the information loss. That said, using our method we were able to achive satisfactory accuracy even with high numbers of quasi-identifiers. This enables providing a higher level of privacy without suffering too large an accuracy loss. 

\subsection{Defending against inference attacks}
\label{sub:def}

\begin{table*}[ht]
	\centering
	\small
	\begin{tabular}{p{5.6cm}||p{1.3cm}|p{1.3cm}|p{1.3cm}|p{1.3cm}|p{1.1cm}}
		Dataset, model \& mitigation & Train \newline accuracy & Test \newline accuracy & Attack accuracy & Precision & Recall \\
		\hline \hline
		Adult, RF & 0.98 &  0.83 & 0.58 & 0.56 & 0.78 \\
		\hline
		Adult, RF, k=50 & 0.85 &  0.83 & 0.51 & 0.51 & 0.33 \\
		\hline
		Adult, RF, k=100 & 0.85 &  \cellcolor{blue!25}0.81 & 0.5 & 0.5 & 0.76 \\
		\hline
		Adult, RF, epsilon=10 & 0.82 &  0.81 & 0.53 & 0.52 & 0.95 \\
		\hline
		Adult, RF, epsilon=1 & 0.8 &  \cellcolor{blue!25}0.79 & 0.5 & 0.52 & 0.13 \\
		\hline \hline
		Adult, NN & 0.96 &  0.89 & 0.53 & 0.52 & 0.6 \\
		\hline
		Adult, NN, k=25 & 0.84 &  0.82 & 0.5 & 0.5 & 0.71 \\
		\hline
		Adult, NN, k=50 & 0.84 &  0.81 & 0.5 & 0.5 & 0.32 \\
		\hline
		Adult, NN, epsilon=2.11, delta=e-05 & 0.83 &  0.83 & 0.5 & 0.5 & 0.14 \\
		\hline \hline
		Loan, RF & 1.0 &  0.92 & 0.67 & 0.65 & 0.71 \\
		\hline
		Loan, RF, k=100 & 1.0 &  \cellcolor{blue!25}0.92 & 0.51 & 0.51 & 0.29 \\
		\hline
		Loan, RF, epsilon=10 & 0.82 &  \cellcolor{blue!25}0.81 & 0.5 & 0.66 & 0.0094 \\
		\hline
		Loan, RF, epsilon=1 & 0.82 &  0.8 & 0.5 & 0.5 & 0.18 \\
		\hline \hline
		Loan, NN & 0.95 &  0.95 & 0.54 & 0.53 & 0.67 \\
		\hline
		Loan, NN, k=200 & 0.95 &  \cellcolor{blue!25}0.92 & 0.5 & 0.5 & 0.46 \\
		\hline
		Loan, NN, epsilon=3.1, delta=e-05 & 0.87 &  \cellcolor{blue!25}0.87 & 0.5 & 0.5 & 0.47 \\
	\end{tabular}
	\centering
	\caption{Mitigations against membership inference. The first two columns refer to the attacked (target) model and the last three refer to the attack. Precision and recall refer to the positive (member) class.}
	\label{table:mit}
\end{table*}

%\vspace{-2em}

Next, we demonstrate the effect of using our anonymization method to defend against membership inference attacks. We trained both a random forest and neural network model on the Adult and Loan datasets. The random forest has the same hyperparameters as in the previous experiment. The neural network architecture consists of three fully-connected layers of sizes 1024, 512 and 256, tanh activation, and Adam optimizer with learning rate 0.0001. We employed a more complex architecture this time, along with tanh activation, to make it easier to successfully apply membership inference to the model. Each model's initial accuracy is documented in Table \ref{table:mit}.

The attack model consisted of an architecture similar to the one used by Nasr et al. \cite{reg} but slightly shallower and with smaller layers (see Appendix \ref{app:attack}). Also similar to the Nasr paper, we employed a \textbf{strong} attack, that assumes knowledge of (part of) the training data used to train the model. 

After applying the attack to the original models, we then trained each of the models on different anonymized versions of the datasets, using different k values. For the Adult dataset we used all features as QI and for the Loan dataset we used the same set of 18 features as in previous experiments. Those results are presented in Table \ref{table:mit}. It is clear from these results that the attack can be mostly prevented using our anonymization method, with a very low accuracy cost. 

For comparison we also applied differential privacy (DP) as a possible mitigation against the membership attack. For the random forest model we used the implementation of Fletcher and Islam\footnote{https://github.com/sam-fletcher/Smooth\_Random\_Trees} \cite{rf}. For the neural network we used the torch-dp library\footnote{https://github.com/facebookresearch/pytorch-dp}, whose implementation is based on the principles of Abadi et al. \cite{dp}. The results achieved using differential privacy were comparable to our anonymization results in terms of the achieved model accuracy versus attack accuracy. In some cases, DP  achieved slightly higher model accuracy for the same attack mitigation level, but in most cases the DP implementation did slightly worse. With the Loan dataset there was a pretty significant 5-11\% test accuracy gap in favor of our method. Those results can also be found in Table \ref{table:mit}, the cells highlighted in blue indicating the cases where our method outperformed DP in terms of the privacy-accuracy tradeoff. 

It is worth noting that the difference in performance overhead when using differential privacy was very noticeable, as well as the added complexity of having to use different implementations for the different model types and ML frameworks. In the case of our anonymization approach, the same method can be applied regardless of the ML model, framework and implementation details.

We also tested the efficiency of using our anonymization method to defend against attribute inference attacks, such as the one described in \cite{confidence}. For this evaluation we used the Nursery dataset\footnote{https://archive.ics.uci.edu/ml/datasets/nursery}, and the `social' feature as the attacked feature. Table \ref{table:mit2} shows the effectiveness of the attack on a decision tree model, with and without applying anonymization to the dataset. These results show that our method can also be effective against attribute inference.

It is important to note that the anonymization-based method does not work well for highly dimensional datasets, especially with a high degree of uniqueness. For example, applying this method to the Purchase100 dataset from \cite{membership}, we were not able to achieve more than 24\% model accuracy, even using the smallest possible value of k (k=2). In contrast, the DP implementation was able to achieve up to 42\% accuracy (still significantly lower than the original 66\%).

\begin{table}
	\centering
	\small
	\begin{tabular}{p{5cm}||p{2cm}|p{2cm}|p{1.3cm}|p{1cm}}
		Dataset, model \& mitigation & Test \newline accuracy & Attack \newline accuracy & Precision & Recall \\
		\hline \hline
		Nursery, DT & 0.68 &  \cellcolor{blue!25}0.61 & \cellcolor{blue!25}0.41 & \cellcolor{blue!25}0.32 \\
		\hline
		Nursery, DT, k=100 & 0.74 &  \cellcolor{blue!25}0.56 & \cellcolor{blue!25}0.31 & \cellcolor{blue!25}0.21 \\
	\end{tabular}
	\centering
	\caption{Mitigation against attribute inference. The first column refers to the attacked model and the last three refer to the attack. Precision and recall refer to the `problematic' value of the social feature. All features were used as quasi-identifiers.}
	\label{table:mit2}
\end{table}

\vspace{-2em}

\section{Conclusions and future work}
\label{sec:con}
We presented a novel machine learning model anonymization method that applies a model-specific accuracy-guided anonymization to the training dataset, then retrains the model on the anonymized data to yield an anonymized model. This anonymized model no longer contains any personal data, and can be freely used or shared with third parties. 

The accuracy-guided anonymization method outperforms state-of-the-art anonymization techniques in terms of the achieved utility, as measured by the resulting model's accuracy. It also enables applying anonymization in previously unattainable situations: when employing complex machine learning models, using large values of k and large numbers of quasi-identifiers. This provides stronger privacy than previously possible, without suffering from great accuracy loss, making anonymous machine learning a feasible option.

In addition, this method is able to achieve a similar effect in preventing membership inference attacks as alternative approaches based on differential privacy, while being much less complex and resource-intensive. It does not assume the availability of similar public datasets, and does not require making changes to the training process. It is thus well suited for scenarios where the private data owner does not have complete control over the training process, or for organizations that employ many different ML models. Moreover, the presented method can defend against other classes of attacks such as attribute inference.

In this work, only the training data underwent anonymization, and the test (runtime) data remained unchanged. It could be interesting to investigate whether applying the same transformations to the test data as to the training data would improve the performance of the model. This could be accomplished by storing those transformations as part of the ML pipeline to be applied on any input data to the model. Moreover, the use of decision trees and knowledge distillation may lend to relatively easily extending this work to support regression tasks in addition to classification. This can be achieved by employing a decision tree regressor as the anonymization model. Finally, we would like to investigate how ML anonymization could potentially be integrated with other related technologies such as federated learning and homomorphic encryption.

%
% ---- Bibliography ----
%
% BibTeX users should specify bibliography style 'splncs04'.
% References will then be sorted and formatted in the correct style.
%
\bibliographystyle{splncs04}
\bibliography{paper}

%\newpage
\appendix

%\appendixpage

%\begin{appendices}

\section{Datasets and Quasi-identifiers}
\label{app:QI}
Table \ref{table:data} describes the datasets used for evaluation. For each dataset the table shows: the number of records, the number of features in the downloaded data, the number of features used (before applying feature selection as part of the training process), the number of categorical features, and the label feature. Table \ref{table:QI} presents the attributes used as quasi-identifiers in the different runs.

\begin{table}[ht]
	\small
	\begin{tabular}{p{0.9cm}||p{1.1cm}|p{2cm}|p{2cm}|p{1cm}|p{4cm}}
		Name & records & total attrs & attrs used & categ & label \\
		\hline \hline
		Adult & 48842 &  14 & 12 & 7 & income \\
		\hline
		Loan & 421095 &  144 & 43 & 11 & loan status \\
	\end{tabular}
	\caption{Datasets used for evaluation}
	\label{table:data}
\end{table}

\vspace{-4em}

\begin{table}[hbt!]
	\small
	\begin{tabular}{p{1.1cm}||p{0.8cm}|p{12cm}}
		Dataset & \# QI & QI attributes \\
		\hline \hline
		Adult & 12 & age, workclass, education-num, marital-status, occupation, relationship, race, sex, capital-gain, capital-loss, hours-per-week, native-country  \\
		\hline
		Adult & 10 & age, workclass, education-num, marital-status, occupation, relationship, race, sex, hours-per-week, native-country  \\
		\hline
		Adult & 8 & workclass, marital-status, occupation, relationship, race, sex, native-country, education-num  \\
		\hline
		Loan & 18 & emp\_length, home\_ownership, annual\_income, zip\_code, purpose, dti, delinq\_2yrs, inq\_last\_6mths, mths\_since\_last\_delinq, open\_acc, total\_acc, mths\_since\_last\_record, pub\_rec, revol\_bal, revol\_util, hardship\_flag, last\_pymnt\_amnt, installment  \\
	\end{tabular}
	\caption{Quasi-identifiers used for evaluation}
	\label{table:QI}
\end{table}

\vspace{-2em}

\section{Attack model}
\label{app:attack}
\begin{figure}
	\centering
	\includegraphics[width=5cm]{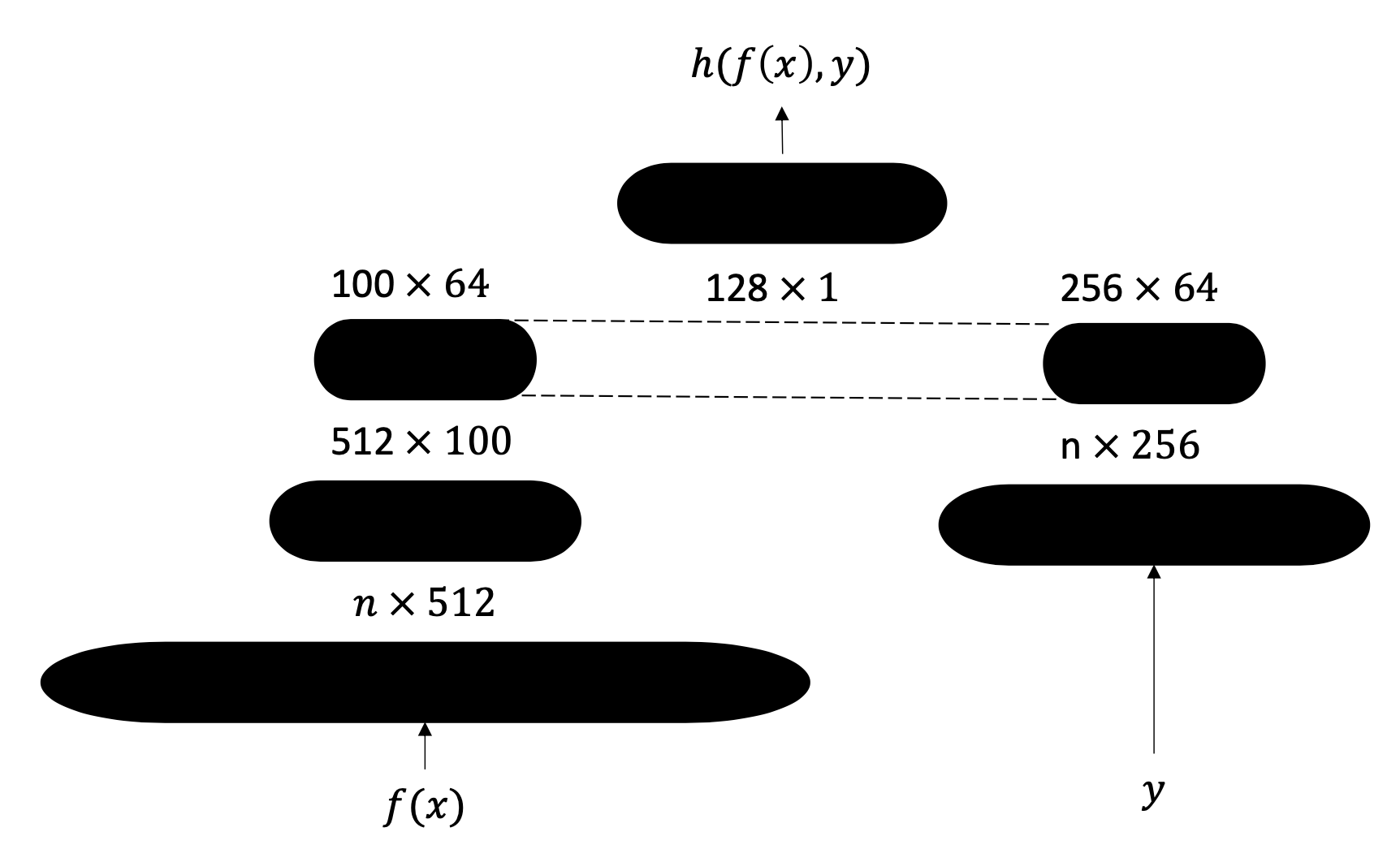}
	\caption{Attack model}
	\label{fig:nn}
\end{figure}

Figure \ref{fig:nn} depicts the attack model employed for membership inference. n represents the number of target classes in the attacked model, f(x) is either the logit (for neural network) or probability (for random forest) of each class, and y is the one-hot encoded true label. This architecture was adapted from \cite{reg} and chosen empirically to yield better accuracy for the tested datasets and models.
We trained the attack model with 50\% members and 50\% non-members.
%In some cases we observed that using a simple random forest as the attack model also worked similarly well.

\vspace{-1.2em}

\end{document}